\documentstyle[11pt,newpasp,twoside,epsf]{article}

\begin{document}

\title{Structural properties of Dark Matter Halos}
\author{Vincenzo Antonuccio-Delogu}
\affil{Laboratory for Computational Astrophysics, Catania Astrophysical
Observatory, Citt\'{a} Universitaria, Via Santa Sofia 78, I-95125 Catania, ITALY}
\author{Ugo Becciani}
\affil{Laboratory for Computational Astrophysics, Catania Astrophysical
Observatory, ITALY}
\author{Antonio Pagliaro}
\affil{Stichting Ruimte Onderzoek Nederland, Sorbonnelaan 2,
3584 CA Utrecht, THE NETHERLANDS}

\begin{abstract}
Using the results of a high mass resolution ($256^{3}$ particles) N-body simulation 
of a cluster-forming region, we study the statistic $\sigma_{v}-M$ for low mass
halos (${\rm M} < 10^{12} {\rm M}_{\sun}$), and we compare it with three 
theoretical models. At the final redshift halos are well
described by the Truncated Isothermal Sphere model recently 
introduced by Shapiro, Iliev \& Raga (1999).
We speculate that this is a consequence of the action of  
tidal fields on galaxy
formation within clusters.

\end{abstract}


\section{Introduction}
The study of the properties of halos produced in N-body 
simulations is of central importance to semianalytical
modelling of galaxy formation and evolution within clusters.
During the past four years, most attention has been paid to one
out of the many possible statistics which could characterize the
halo population, namely the density profile (Navarro, Frenk \&
White 1996; Tormen, Bouchet \& White, 1997; Moore et al., 1999; 
Jing \& Suto, 1999).
However, it is difficult to determine reliably the density profile 
of {\em numerical} dark matter halos containing less than
$\approx 10^{5}$ particles: and in modern high resolution,
parallel N-body cosmological simulations, there are typically
not so many rich halos which are not formed by ``overmerged''
material in the center of clusters. In fact, all the work in the
above mentioned papers where the density profiles are studied
has been performed on halos extracted from some simulations
and ``re-simulated'' at higher mass and spatial resolution. This
allows one to study in detail at most a dozen of halos, but its is
difficult to perform with this technique an analysis spanning a
wide range of the possible halos' parameters range.
On the other hand, numerical simulations typically produce a lot
of halos with $\approx 10^{1}-10^{4}$ particles, which carry a
significant dynamical information having been ``processed'' by
the gravitational field of the region where they form. It could be
wise to try to use this statistical information to try to
discriminate among different models of structure formation.
Unfortunately we do not have a complete physical
understanding of the gravitational instabilities which drive a
halo (and particularly a {\em numerical} halo from an
intrinsically spatially and temporally discrete experiment
as a N-body simulation is) toward a (almost) relaxed state, so
we must adopt models to interpret the results of numerical
simulations. In this contribution we show that, given enough
spatial and mass resolution, and choosing an appropriate
statistics, it is possible to discriminate among different models
of halo formation.

\section{Models of collapse}
We consider three models for gravitational collapse of halos:
Singula Isothermal Sphere (SIS), the
spherically-averaged peak-patch model by Bond \& Myers (1996)
and the Truncated Isothermal Spherical (TIS) model by Shapiro, 
Iliev \& Raga (1999). The physical assumptions of the three
models are significantly different. The SIS model is the
simplest one, but also the most unrealistic: it predicts a
singular density profile decaying $\rho_{SIS}\propto r^{-2}$.
Total mass is infinite and density is everywhere
nonzero (the system is not truncated).\\
The Bond \& Myers model is based on a Montecarlo approach
named by their 
authors ``peak-patch''. The underlying model is that of a collapse
of a homogenous spheroid, for which the equations of motion are
exactly solvable. We consider here the spherically-averaged
quantities computed from this model. The third and last is a
recent class of isothermal, truncated models recently introduced
by Shapiro et al. (1999). The main idea is to consider solutions
of the Jeans' equations for isothermal systems truncated at a
finite radius. This is possible only if the shear gradient terms
in these equations are not identically zero, for instance if an
isotropic pressure term is present.\\
The relationship between 1-D velocity dispersion and mass in
these three models is given by
\begin{equation}
\sigma_{v}=C_{SIS, TIS}\left( M_{12}\right)^{1/3}\left(
1+z_{coll}\right)^{1/2} h^{1/3}
\end{equation}
where: $C_{SIS} = 71.29$ and $C_{TIS} = 104.69$ (in km/sec),
respectively, and $ M_{12}$ is the mass in units of $10^{12}
{\rm M}_{\sun}$. For the Bond \& Myers model the relationship is
slightly different:
\begin{equation}
\sigma_{v}=117.6\left( M_{12}\right)^{0.29}\left(
1+z_{coll}\right)^{1/2} h^{1/3}
\end{equation}
(Bond \& Myers 1996, eq. 4.4). Note that this latter
relationship was obtained from a fit of Montecarlo simulations
for a range of mass much larger than the one we consider in this
contribution (see next section).

\section{Simulation and results}
In order to test the models presented in the previous section, we
will use the results from a N-body simulation we have recently
performed. Initial conditions were picked up from the catalogue
of simulated clusters by van Kampen \& Katgert (1997): we choose
a configuration which would produce a double cluster, so we
could study galaxy formation in a high shear environment. We
then have run the same set of initial conditions with a higher
mass and spatial resolution. The run was performed with
$256^{3}$ particles using a parallel treecode (Becciani,
Antonuccio-Delogu \& Pagliaro, 1996). The softening length was
fixed at 15 $h^{-1}$ kpc (comoving). Although the simulation
produces about a dozen of clusters, we will restrict our
attention to the two major clusters, which are shown in 
Figure~\ref{fig3}\\
In order to study the virialization properties, it is important
to adopt a physically motivated criterion to select the halos.
This is because these properties are traced
by gravitationally bound particles. Simple criteria (like
friends-of-friends) do not distinguish particles on the base of
their gravitational properties, but only on the base of their
relative distance. 
\begin{figure}
\plotfiddle{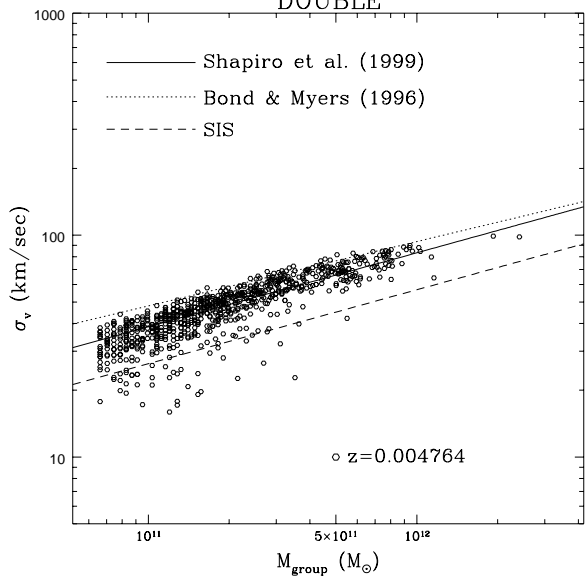}{5cm}{0}{40}{40}{-150}{-100} \label{fig2}
\vspace*{1cm}
\caption{Virialization properties of the region around the
DOUBLE cluster (see text). Gravitationally bound groups were
identified by {\em SKID}, using the same parameters as in the
simulation and a standard percolation length 0.2$r_{av}$. Only
groups containing at least 50 particles were included in the
final catalogue.}
\end{figure}
Here we selected groups using {\em SKID}, a
open source software which produces catalogues of groups
including only gravitationally bound particles (Stadel, Katz \& 
Quinn, 1999). The only serious drawback concerning {\em SKID} is
that it is very slow. For this reason, we restricted our
analysis to a $10^{3} h^{-3} {\rm Mpc}^{3}$ region centered
around the double cluster and a similar region centered around a
single cluster. We denote these two regions as DOUBLE and
SINGLE in the following.
\begin{figure}
\plotfiddle{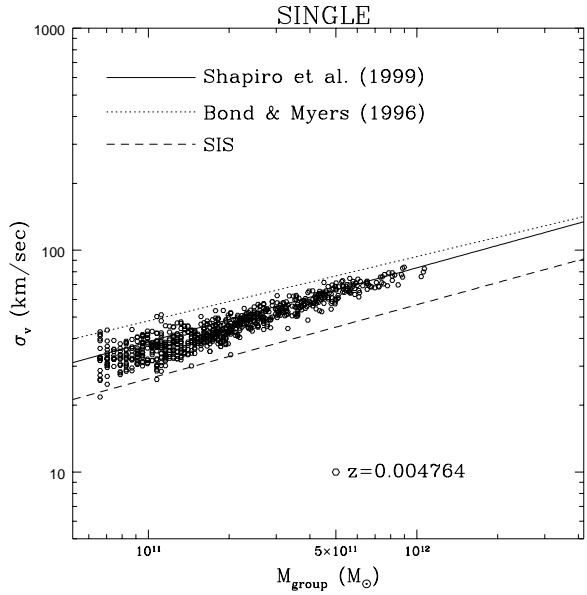}{5cm}{0}{40}{40}{-150}{-100} \label{fig3}
\vspace*{1cm}
\caption{Virialization properties of the region around the
SINGLE cluster (see text). The same parameters as in Figure 2
were adopted for {\em SKID}. }
\end{figure}

DOUBLE and SINGLE contain approximately the same number of
objects. Here we focus our attention on halos with masses in the
range $10^{11}-10^{12} {\rm M}_{\sun}$, i.e. low-mass halos.
These are the smallest objects we can reliably trace
with the mass resolution and softening length we adopted.
Within this mass range
DOUBLE has 827 halos, while SINGLE has 757. From
Figures 1 and 2
we can see that using our halo samples it is possible 
to discriminate among the three models outlined in the
previous section. Both plots show that the TIS model offers a
better description of the equilibrium properties of these halos.
But notice also that a fraction of halos in
DOUBLE has velocity dispersions significantly smaller than the
average. Tidal effects, which are much more pronounced in DOUBLE
than in SINGLE, are responsible for this difference
(Antonuccio-Delogu, Pagliaro \& Becciani, 1999b).

\section{Conclusion}
The fact that the TIS model gives a better fit to the properties of
low-mass virialized halos should not be surprising: it is
possible to show that this reflects the action of the
environmental tidal field on halo formation (Antonuccio-Delogu 
et al., 1999a). However, we have found that the actual prfile
of these halos is different from the Minimum Energy state
suggested by Shapiro et al. (1999), and is rather more
consistent with a tidally limited profile.\\
Before concluding, a few words about the consistency of the
results of our work with the idea of the existence of a
``universal'' density profile. The TIS density profile differs
significantly from either the Navarro et al. (1996) and the
Moore et al. (1998) density profiles, because it flattens in
the central region (i.e. it has a core). At the time we wrote
this contribution, Jing \& Suto (1999) submitted the results of
a series of simulations which suggest a flattening of the inner
2\% (in units of $r/r_{200}$) of the density profile in
galaxy-sized halos. A similar flattening is not observed in
their cluster-sized halos. The issue is then still an open one.\\
All this seems to suggest that statistics based on
virialization properties 
bear a smaller intrinsic uncertainty than the density
profile, and are then more suitable
to characterize the average statistical properties of halo
populations.

\acknowledgments
V.A.-D. would like to thank E. van Kampen for having supplied
the initial parameters from his catalogue of simulated clusters.

\newpage
\begin{figure}
\plotone{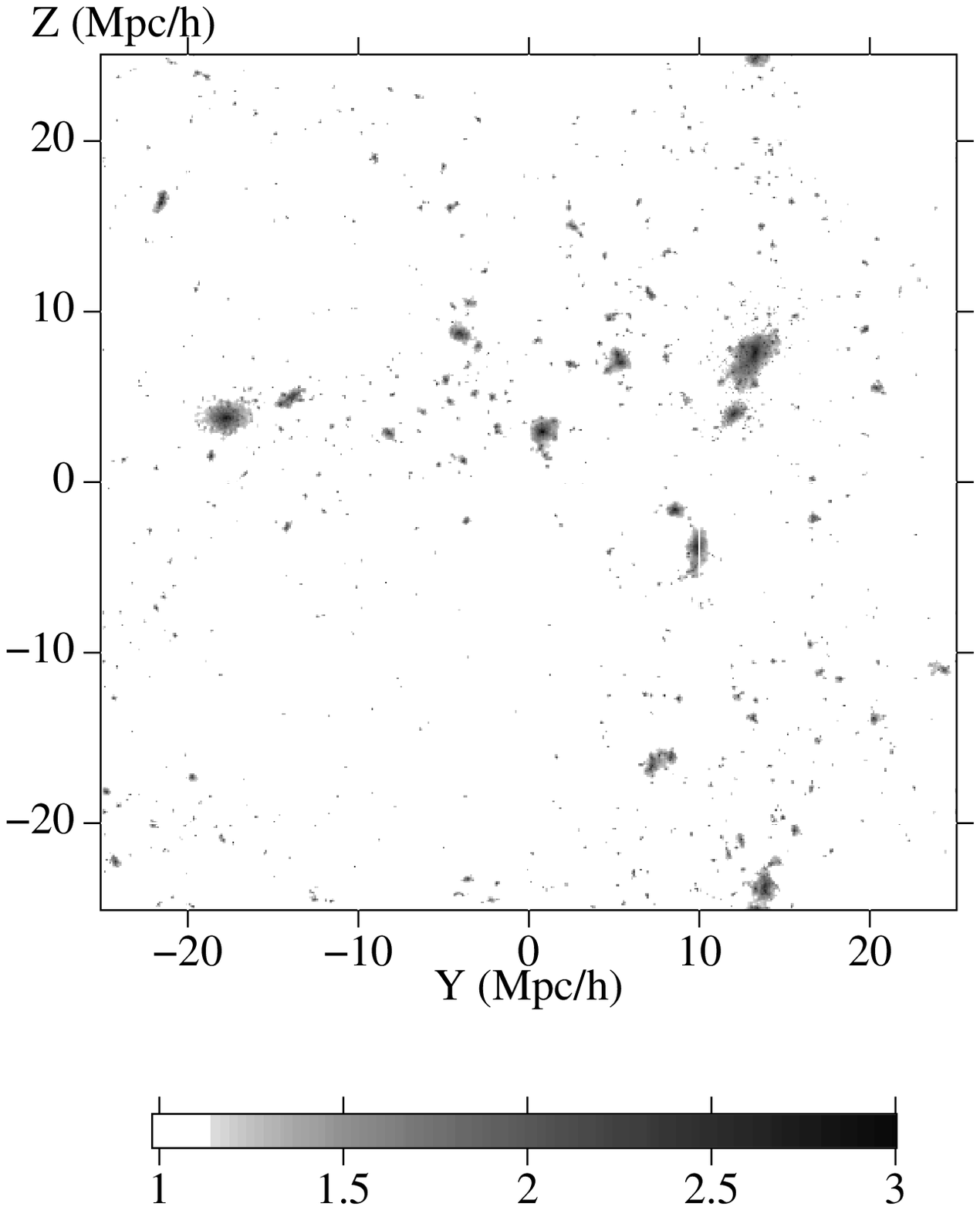} \label{fig1}
\caption{Final output of the simulation. The two regions from
which low-mass galactic sized halos were extracted are
centered around the two most massive clusters at (13,8) and
(-18,3).}
\end{figure}

\end{document}